\documentclass{iopart}

\usepackage{iopams}
\usepackage{setstack}
\usepackage{graphicx}

\begin{document}

\title{Unbiased estimation of an angular power spectrum}

\author{G. Polenta\dag, D. Marinucci\ddag, A. Balbi\S, P. de
  Bernardis\dag, E. Hivon$\|$, S. Masi\dag, P. Natoli\S, N. Vittorio\S}
\address{\dag\ Dipartimento di Fisica, Universita' di Roma ``La Sapienza''}
\address{\ddag\ Dipartimento di Matematica, Universita' di Roma ``Tor
  Vergata''}
\address{\S\ Dipartimento di Fisica, Universita' di Roma ``Tor
  Vergata''}
\address{$\|$\ IPAC,California Institute of Technology}
\ead{\mailto{gianluca.polenta@roma1.infn.it}}

\begin{abstract}
We discuss the derivation of the analytic properties of the cross-power spectrum 
estimator from multi-detector CMB anisotropy maps.
The method is computationally convenient and it provides unbiased 
estimates under very broad
assumptions.
We also propose a new procedure for testing for the presence of
residual bias due to inappropriate noise subtraction in pseudo-$C_{\ell}$ 
estimates. We
derive the analytic behavior of this procedure under the null hypothesis, 
and use Monte Carlo simulations to investigate its efficiency properties, 
which appear very promising. For instance, for full sky maps with isotropic white
noise, the test is able to identify an error of 1\% on the noise
amplitude estimate.
\end{abstract}

\pacs{95.75Pq, 98.80Es, 02.50Ng, 02.50Tt}

\section{Introduction}

The cosmic microwave background (CMB) provides one of the most powerful
ways of investigating the physics of the early Universe. The main CMB
observable is the angular power spectrum of temperature anisotropy, which
encodes a large amount of cosmological information. In the last decade,
important advances in the measurement of the CMB angular power
spectrum took place; this resulted in relevant progress in our understanding of physical
cosmology. CMB temperature anisotropies were first detected by the COBE
satellite in 1992 \cite{smoot92}. 
This discovery fuelled a period of
intensive experimental activity, focused on measuring the CMB power spectrum
on a large range of angular scales. A major breakthrough was made in the
past few years, when the MAXIMA \cite{hanany00} and BOOMERanG \cite{debe00}
balloon-borne experiments independently produced
the first high-resolution maps of the CMB, allowing a clear measurement of a
peak in the power spectrum, as expected from theoretical models and previously
detected by the ground based experiment TOCO \cite{miller99}. 
Since then,
many other experiments have confirmed and improved on these results: DASI\cite{dasi02}, 
BOOMERanG-B98 \cite{nett02,debe02,ruhl03}, BOOMERanG-B03 \cite{masi05,
jones05, fp05, montroy05, cmt05}, VSA \cite{vsa03}, Archeops \cite{benoit03}, 
CBI \cite{cbi02}, ACBAR\cite{acbar02}, BEAST \cite{beast03}. Most notably, 
the NASA satellite mission
WMAP, whose first year data were released in February 2003 
(\cite{wmap03} and references therein), provided the first
high-resolution, full sky, 
multi-frequency CMB maps, and a determination of the angular power spectrum with unprecedented
accuracy on a large range of angular scales. Much larger and more accurate data 
sets are expected in the years to come from ESA's Planck satellite.

In this paper, we shall concentrate on extracting the CMB power spectrum
from full sky maps with foregrounds removed. We shall focus
mainly on techniques for dealing with noise subtraction. In principle, and for
Gaussian maps, noise subtraction can be performed
by implementing maximum likelihood estimates.
It is well known, through \cite{bjk98,borrill99}, 
that maximum likelihood estimates require for their
implementations a number of operations that scales as $N_{pix}^{3},$ 
$N_{pix} $ denoting the number of pixels in the map. For current experiments, 
$N_{pix}$ ranges from several hundred thousands to a few millions, and thus the
implementation of these procedures is beyond computer power for
the near future. Many different methods have been proposed for producing
computationally feasible estimates; here we just mention a few of them, and
we refer the reader to \cite{efst03} for a more complete discussion on their merits. Some
authors have introduced special assumptions on the noise properties and
symmetry of the sky coverage, to make likelihood estimates 
feasible; see, for instance, \cite{oh99,wgh01,wh01,ch01}. 
Reference~\cite{sza01} adopted an entirely different strategy, extracting the power spectrum from
the 2-point correlation function of the map. Others have used 
estimators based on pseudo-$C_l$ statistics and Monte Carlo
techniques \cite{efh01,balbi02}, or based on Gabor transforms \cite{fh02}.
For multi-detector experiments, an elegant method, based on spectral matching to estimate
jointly the angular power spectrum of the signal and of the noise, was
proposed in \cite{dela}. 
Pseudo-$C_{\ell}$ estimators were adopted by the WMAP team \cite{wmapps}, which used
the cross-power spectrum estimator and discussed the best combination of the
cross-power spectrum obtained from single couples of receivers.

Our purpose in this paper is to derive some analytic results on the
cross-power spectrum estimator, to perform a comparison with standard
pseudo-$C_{\ell}$ estimators, and to propose some testing procedures on the
assumption that any noise bias has been properly removed, which 
is clearly a crucial step in any estimation approach. We shall also present some 
Monte Carlo evidence on the performance of the methods that we advocate. 
The plan of this paper is as follows. In Section \ref{sec:ps} we derive the analytic
properties for the cross-power spectrum estimator and we compare them with
equivalent results on standard pseudo-$C_{\ell}$ estimators. In Section 
\ref{sec:haus} we propose a procedure (the Hausman test) for verifying appropriate noise
subtraction in pseudo-$C_{\ell}$ estimators, and we derive its analytic properties.
In Section \ref{sec:mc} we validate our results by using Monte Carlo
simulations, which are also used to test the power of our procedure in the
presence of noise which has not been completely removed. In Section 
\ref{sec:con} we review our results and discuss directions for future research.

\section{Power spectrum estimators}

\label{sec:ps} The CMB temperature fluctuations $\frac{\Delta T}{T}(\theta
,\phi )$ can be decomposed into spherical harmonic coefficients

\begin{equation}
a_{\ell m}=\int_{\Omega} \frac{\Delta T}{T}(\theta, \phi) \overline{Y}%
^m_{\ell}(\theta, \phi)d\Omega \mbox{ .}
\end{equation}

If the CMB fluctuations are Gaussian distributed and statistically homogeneug, as suggested by the
latest experimental results (see for instance \cite{pol, wmapng, maxng}), then each $a_{\ell
m}$ is an independent Gaussian complex variable with 
\begin{eqnarray}
&\langle a_{\ell m}\rangle = 0 & \\
& \langle a_{\ell m}\overline{a_{\ell ^{^{\prime }}m^{^{\prime }}}}\rangle
= \delta _{\ell \ell ^{^{\prime }}}\delta _{mm^{^{\prime }}}C_{\ell } &
\end{eqnarray}%
and all the statistical information is contained in the power spectrum $%
C_{\ell }$.

In the following we describe two procedures for estimating the CMB angular
power spectrum: the standard pseudo-$C_{\ell}$ estimator, sometimes labelled the
auto-power spectrum \cite{wmapps}, and the cross-power spectrum.
As a first step, we shall assume handling of full sky maps with
isotropic, not necessarily white noise.

\subsection{The standard pseudo-$C_{\ell}$ estimator}

Pseudo-$C_{\ell}$ estimators are very useful in computing the power spectrum because
they are fast enough to be used on large data sets such as WMAP and Planck. The
standard pseudo-$C_{\ell}$ estimator has been thoroughly investigated in the
literature, taking also into account some important features of realistic
experiments such as partial sky coverage and systematic effects \cite{efh01}. 
The starting point is the raw pseudo-power spectrum $\widehat{C}%
_{\ell }^R$ defined as 
\begin{equation}
\widehat{C}_{\ell }^R=\frac{1}{2\ell +1}\sum_{m=-\ell }^{\ell }d_{\ell m}%
\overline{d_{\ell m}}
\end{equation}%
where $d_{\ell m}$ are the spherical harmonic coefficients of the map.

In the absence of noise and for a full sky CMB map, 
$d_{\ell m}=a_{\ell m}$ and $\widehat{C}_{\ell}^R$
is an unbiased
estimator of $C_{\ell }$ (the angular power spectrum of the signal) with
mean equal to $C_{\ell }$ and variance equal to $2C_{\ell }/2\ell +1$; also, 
$(2\ell +1)\widehat{C}_{\ell }^R/C_{\ell }$ is a $\chi _{\nu }^{2}-$%
distributed variable with $\nu =2\ell +1$ degrees of freedom.

In the presence of noise, it is not difficult to see that this estimator is
biased.
If we assume, as usual, that noise is independent from the signal, we have
\begin{equation} \label{eq:dl}
  d_{\ell m}=a_{\ell m}+a_{\ell m}^N
\end{equation}
and 
\begin{equation}
  \langle \widehat{C}_{\ell }^R \rangle =C_{\ell }+C_{\ell }^{N}
  \mbox{ .}
\end{equation}%
Now the common assumption is to take $C_{\ell }^{N}$ as determined \emph{a
priori}, for instance by Monte Carlo simulations 
and measurements of the properties of the detectors; we shall discuss later how to
test the validity of this assumption and/or make it weaker. Under these
circumstances, the power spectrum estimator is naturally defined as
\begin{equation}
\widehat{C}_{\ell }= \widehat{C}_{\ell }^R -C_{\ell }^{N} =
\frac{1}{2\ell +1}\sum_{m=-\ell }^{\ell }d_{\ell m}%
\overline{d_{\ell m}}-C_{\ell }^{N} \mbox{ .} \label{basest} 
\end{equation}%
Of course, if the estimate of the noise power spectrum $C_{\ell }^{N}$ is
not correct, the estimator will be biased. For a multi-channel experiment,
we generalize equation~\ref{basest} by averaging the maps from each detector and
then computing the power spectrum of the resulting map. A more
sophisticated approach would be to use weighted averages, with weights
inversely proportional to the variance of each detector, but we shall not
pursue this idea for the sake of brevity. In view of equations~\ref{eq:dl} and~\ref{basest},
 in the presence of $k$ channels with
uncorrelated noises we can write 
\begin{equation}
\widehat{C}_{\ell }=\frac{1}{2\ell +1}\sum_{m=-\ell }^{\ell }\left[ \left|
a_{\ell m}+\frac{1}{k}\sum_{i=1}^{k}a_{\ell m}^{N_{i}}\right| ^{2}-\frac{%
1}{k^{2}}\sum_{i=1}^{k}\langle \widehat{C}_{\ell }^{N_{i}}\rangle \right]
\end{equation}%
where $i$ is the detector index and $a_{\ell m}^{N_i}$ are the noise
spherical harmonics coefficients. 
Assuming that our noise estimation is
correct, we obtain for the expected value and the variance 
\begin{equation}
<\widehat{C}_{\ell }>=C_{\ell }
\end{equation}%
and 
\begin{equation}
\fl Var\left\{ \widehat{C}_{\ell }\right\} =\frac{2}{2\ell +1}\left\{ C_{\ell
}^{2}+\frac{2}{k^{2}}\sum_{i=1}^{k}C_{\ell }C_{\ell }^{N_{i}}+\frac{1}{k^{4}}%
\left[ \sum_{i=1}^{k}\sum_{j=1}^{k}C_{\ell }^{N_{i}}C_{\ell }^{N_{j}}\right]
\right\} \mbox{ .}\label{mastervar}
\end{equation}%
It should be noted that in equation~\ref{mastervar} the value of $C_{\ell }^{N}$ is
taken as fixed, and in this sense we are underestimating the variance by
neglecting the additional uncertainty due to the estimation of the noise
properties.

\subsection{The cross-power spectrum}

The pseudo-$C_{\ell}$ estimator presented in the previous subsection is
computationally very fast and simple to use, but it is prone to bias if
noise has not been appropriately removed. It is thus natural to look for more
robust alternatives, yielding unbiased estimates even in the presence of
noise with an unknown angular power spectrum. For this purpose, we now focus
on the cross-power spectrum, which is defined, for any given couple of
channels $(i,j)$, as 
\begin{equation}
\widetilde{C}_{\ell }^{ij}=\frac{1}{2\ell +1}\sum_{m=-\ell }^{\ell }d_{\ell
m}^{i}\overline{d}_{\ell m}^{j}  \mbox{ .} \label{eq:def}
\end{equation}%
It iss easy to show that 
\begin{equation}
\langle \widetilde{C}_{\ell }^{ij}\rangle =C_{\ell }
\end{equation}%
and 
\begin{equation}
Var\left\{ \widetilde{C}_{\ell }^{ij}\right\} =\frac{2}{2\ell +1}\left\{ C{%
_{\ell }^{2}}+\frac{C{_{\ell }}}{2}(C_{\ell }^{N_{i}}+C_{\ell }^{N_{j}})+%
\frac{C_{\ell }^{N_{i}}C_{\ell }^{N_{j}}}{2}\right\} \mbox{ .}
\label{eq:var}
\end{equation}%
For the details of the calculations see the appendix. Let us now consider
the most general case with $k$ detectors; this means that we can construct $%
k(k-1)/2$ different couples of channels. For each of them we can calculate
the cross-power spectrum and then take the average; thus the cross power
spectrum becomes 
\begin{equation}
\widetilde{C}_{\ell }=\frac{2}{k(k-1)}\sum_{i=1}^{k-1}\sum_{j=i+1}^{k}%
\widetilde{C}_{\ell }^{ij} \mbox{ .}%
\end{equation}%
Again, the resulting estimator is clearly unbiased, $\langle \widetilde{C}%
_{\ell }\rangle =C_{\ell }.$ Its covariance is given by 
\begin{eqnarray}
\fl Var\left\{ \widetilde{C}_{\ell }\right\} = Var\left\{ \frac{2}{k(k-1)}
\sum_{i=1}^{k-1}\sum_{j=i+1}^{k}\widetilde{C}_{\ell }^{ij}\right\}
\nonumber \\
 \lo{=}\frac{4}{k^{2}(k-1)^{2}}\left\{
\sum_{i=1}^{k-1}\sum_{j=i+1}^{k}Var\left\{ \widetilde{C}_{\ell
}^{ij}\right\} \right\} \\
+\frac{4}{k^{2}(k-1)^{2}}\left[ 2Cov\left\{ \widetilde{C}_{\ell }^{12},
\widetilde{C}_{\ell }^{13}\right\} +2Cov\left\{ \widetilde{C}_{\ell }^{12},
\widetilde{C}_{\ell }^{14}\right\} +...\right] \mbox{ .} \nonumber
\label{eq:vartot}
\end{eqnarray}%
In order to evaluate this quantity, the first step is to consider the
covariances among different pairs $(i,j),(h,k).$ For $k$ channels we can
construct $c=k(k-1)/2$ different couples and $t=c(c-1)/2$ covariance terms,
which are 
\begin{equation}
\fl
Cov\left\{ \widetilde{C}_{\ell }^{ij},\widetilde{C}_{\ell }^{hk}\right\} =
\left\{
\begin{array}{ll}
\frac{2}{2\ell+1} C_{\ell }^{2} & \mbox{ for }h\neq i,j\mbox{ and
}k\neq i,j \\ 
\frac{2}{2\ell+1}\left\{ C_{\ell }^{2}+\frac{1}{2}C_{\ell
  }C_{\ell}^{N_{i}}\right\} & \mbox{for }h=i\mbox{ or }j\mbox{ and
}k\neq i,j \mbox{ .}
\end{array}
\right.
\label{sinter}
\end{equation}

The next step is to consider how many times we have the $C_{\ell }C_{\ell
}^{N_{i}}/2$ term, for each $i=1,...,k$. This term appears when one of the
two index of a couple is equal to one of the two index of another couple.
This leaves $(k-1)$ possible values for the second index in the first
couple, and $(k-2)$ possible values for the second index in the second
couple; finally we have a factor $1/2$ to take into account symmetries, that
is, the fact that $Cov\left\{ \widetilde{{C}}_{\ell }^{ij},\widetilde{{C}}%
_{\ell }^{hk}\right\} =Cov\left\{ \widetilde{{C}}_{\ell }^{hk},\widetilde{{C}%
}_{\ell }^{ij}\right\} $ (equivalently, we could drop the factor $2$ which
multiplies the covariance terms in equation~\ref{eq:vartot}). The result is that
the single term $C_{\ell }C_{\ell }^{N_{i}}/2$ appears $(k-1)(k-2)/2$
times.

By plugging in equation~\ref{sinter} into equation~\ref{eq:vartot}, we obtain

\begin{equation}
\fl Var\left\{ \widetilde{C}_{\ell }\right\} =\frac{2}{2\ell +1}\left\{ C_{\ell
}^{2}+\frac{2}{k^{2}}C_{\ell }\sum_{i=1}^{k}C_{\ell }^{N_{i}}+\frac{2}{%
k^{2}(k-1)^{2}}\sum_{i=1}^{k-1}\sum_{j=i+1}^{k}C_{\ell }^{N_{i}}C_{\ell
}^{N_{j}}\right\}  \mbox{ .} \label{eq:final}
\end{equation}

It can be verified that for $k=2$, equation~\ref{eq:vartot} reduces to equation~\ref%
{eq:var}. It is interesting to compare this result with the variance of the
classic pseudo-$C_{\ell}$ estimator. We can write immediately

\begin{equation}
\fl Var\left\{ \widetilde{C}_{\ell }\right\} -Var\left\{ \widehat{C}_{\ell
}\right\} =\frac{2}{2\ell +1}\left\{ -\frac{1}{k^{4}}\sum_{i=1}^{k}(C_{\ell
}^{N_{i}})^{2}+\frac{4k-2}{k^{4}(k-1)^{2}}\sum_{i=1}^{k-1}%
\sum_{j=i+1}^{k}C_{\ell }^{N_{i}}C_{\ell }^{N_{j}}\right\} \mbox{ .}%
\label{eq:vardiff}
\end{equation}

Considering the case where $C_{\ell }^{N_{i}}=C_{\ell }^{N}$ for all the
channels, 
we obtain 
\begin{equation}
Var\left\{ \widetilde{C}_{\ell }\right\} -Var\left\{ \widehat{C}_{\ell
}\right\} =\frac{2}{2\ell +1}\left\{ \frac{1}{k^{2}(k-1)}(C_{\ell
}^{N})^{2}\right\} \mbox{ .}
\end{equation}%
Hence, if noise has the same power spectrum over all channels, then the
standard estimator is always more efficient, although clearly the
difference between the two estimators becomes asymptotically negligible as the
number of detectors grows (it scales as $k^{-3})$. 

We have thus shown that the cross-power spectrum estimator provides a robust
alternative to the classical pseudo-$C_{\ell}$ procedure, in that it does not
require any {\it a priori} knowledge of the noise power spectrum.
We shall argue that cross-power spectrum estimates can be
extremely useful even if different procedures are undertaken to estimate the
angular power spectrum; indeed, in the next section we discuss how to test
the assumption that noise has been appropriately removed from the data from a
multi-channel experiment.

\section{The Hausman test}

\label{sec:haus}

In the previous section, we compared the relative efficiency of the two
estimators $\widehat{C}_{\ell }$,$\widetilde{C}_{\ell },$ in the case
where the bias
term in $\widehat{C}_{\ell }$ had been effectively removed. In this section
we propose a testing procedure to verify the latter assumption. Consider the
random variable $G_{\ell }=\left\{ \widehat{C}_{\ell }-\widetilde{C}_{\ell
}\right\} ;$ if $\widehat{C}_{\ell }$ is unbiased, then it is
immediate that $G_{\ell }$ has mean zero, 
with variance 
\begin{equation}
Var\left\{ \widehat{C}_{\ell }-\widetilde{C}_{\ell }\right\} =Var\left\{ 
\widehat{C}_{\ell }\right\} +Var\left\{ \widetilde{C}_{\ell }\right\}
-2Cov\left\{ \widehat{C}_{\ell },\widetilde{C}_{\ell }\right\} ,
\end{equation}%
where 
\begin{equation}
Cov\left\{ \widehat{C}_{\ell },\widetilde{C}_{\ell }\right\} =\frac{2}{k(k-1)%
}\sum_{i=1}^{k-1}\sum_{j=i+1}^{k}Cov\left\{ \widehat{C}_{\ell },\widetilde{C}%
_{\ell }^{ij}\right\} \mbox{ .} \label{eq:hauscov}
\end{equation}%
In the appendix we show that, for a single couple $(i,j),$ we have 
\begin{equation}
\fl Cov\left\{ \widehat{C}_{\ell },\widetilde{C}_{\ell }^{ij}\right\} =\frac{2}{%
2\ell +1}\left\{ C_{\ell }^{2}+\frac{C_{\ell }}{k}\left( C_{\ell
}^{N_{i}}+C_{\ell }^{N_{j}}\right) +\frac{1}{k^{2}}C_{\ell }^{N_{i}}C_{\ell
}^{N_{j}}\right\}  \mbox{ .} \label{bef}
\end{equation}%
Now we use equation~\ref{bef} in equation~\ref{eq:hauscov} and we obtain 
\begin{equation}
\fl Cov\left\{ \widehat{C}_{\ell },\widetilde{C}_{\ell }\right\} =\frac{2}{2\ell
+1}\left\{ C_{\ell }^{2}+\frac{2C_{\ell }}{k^{2}}\sum_{i=1}^{k}C_{\ell
}^{N_{i}}+\frac{2}{k^{3}(k-1)}\sum_{i=1}^{k-1}\sum_{j=i+1}^{k}C_{\ell
}^{N_{i}}C_{\ell }^{N_{j}}\right\}  \mbox{ .}
\end{equation}%
Therefore 
\begin{equation}
\fl Var\left\{ \widehat{C}_{\ell }-\widetilde{C}_{\ell }\right\} =\frac{2}{2\ell
+1}\left\{ \frac{1}{k^{4}}\sum_{i=1}^{k}(C_{\ell }^{N_{i}})^{2}+\frac{2}{%
k^{4}(k-1)^{2}}\sum_{i=1}^{k-1}\sum_{j=i+1}^{k}C_{\ell }^{N_{i}}C_{\ell
}^{N_{j}}\right\} \mbox{ .}  \label{eq:hausvar}
\end{equation}%
The special case $C_{\ell }^{N_{1}}=...=C_{\ell }^{N_{k}}$ gives 
\begin{equation}
Var\left\{ \widehat{C}_{\ell }-\widetilde{C}_{\ell }\right\} =\frac{2}{2\ell
+1}\frac{1}{k^{2}(k-1)}(C_{\ell }^{N})^{2} \mbox{ .}
\end{equation}%
Thus, for a fixed $\ell $ we can suggest the statistic 
\begin{equation}
\fl H_{\ell }=\left( \ell +\frac{1}{2}\right) ^{1/2}k^{2}\left[
\sum_{i=1}^{k}(C_{\ell }^{N_{i}})^{2}+\frac{2}{(k-1)^{2}}\sum_{i=1}^{k-1}%
\sum_{j=i+1}^{k}C_{\ell }^{N_{i}}C_{\ell }^{N_{j}}\right] ^{-1/2}\left\{ 
\widehat{C}_{\ell }-\widetilde{C}_{\ell }\right\} 
\end{equation}%
as a feasible test for the presence of bias in $\widehat{C}_{\ell }.$ By a
standard central limit theorem, we obtain that%
\begin{equation}
H_{\ell} \rightarrow ^{d} N(0,1) \mbox{ as } \ell \rightarrow \infty
\end{equation}%
where $\rightarrow ^{d}$ denotes convergence in distribution and $N(0,1)$
represents a standard Gaussian random variable. In words, for reasonably
large $\ell $ the distribution of $H_{\ell }$ is very well approximated by a
Gaussian, provided that $\widehat{C}_{\ell }$ is actually unbiased; on the other
hand, if this is not the case the expected value of $H_{\ell }$ will be
non-zero. This observation suggests many possible tests for bias, using for
instance the chi-square statistic $H_{\ell }^{2}$ (a value of $H_{\ell }^{2}$
larger than 3.84, the chi-square quantile at 95\%, would suggest that bias
has not been removed at that confidence level). In practice, however, we
have to focus on many different multipoles, $\ell =1,...,L,$ where $L$
depends on the resolution of the experiment and its signal to noise
properties. It is clearly not enough to consider the whole sequence $\left\{
H_{\ell }\right\} _{\ell =1,2,...,L},$ and check for the values above the
threshold, as this does no longer correspond to the 95\% confidence level
(it is obvious that, if $P(H_{\ell }^{2}>3.84)\simeq 5\%,$ then $%
P(\max_{\ell =1,...,L}H_{\ell }^{2}>3.84)>>5\%,$ the exact value being
difficult to determine). 
To combine the information
over different multipoles into a single statistic in a rigorous manner, we suggest the process 
\begin{equation}
B_{L}(r)=\frac{1}{\sqrt{L}}\sum_{\ell =1}^{[Lr]}H_{\ell } , r\in
\lbrack 0,1]
\end{equation}%
where $[.]$ denotes integer part. Of course, other related proposals
could be considered; for instance we might focus on weighted versions
of $B_{L}(r)$, to highlight the contribution from low multipoles, where
it is well known that there are problems with non-maximum likelihood
estimators. 
This modification, however,
would not alter the substance of the discussion that follows.

We note first that $B_{L}(r)$ has mean zero; indeed, 
\begin{equation}
<B_{L}(r)>=\frac{1}{\sqrt{L}}\sum_{\ell =1}^{[Lr]}<H_{\ell }>=0 \mbox{ .}
\end{equation}%
Also, for any $r_{1}\leq r_{2},$ as $L\rightarrow \infty $, 
\begin{equation}
\fl Cov\left\{ B_{L}(r_{1}),B_{L}(r_{2})\right\} =\frac{1}{L}\sum_{\ell
=1}^{[Lr_{1}]}\sum_{\ell =1}^{[Lr_{2}]}<H_{\ell }H_{\ell }>=\frac{1}{L}%
\sum_{\ell =1}^{[Lr_{1}]}<H_{\ell }^{2}>\rightarrow r_{1} \mbox{ .}
\end{equation}%
As $r$ varies in $[0,1]$, $B_{L}(r)$ can be viewed as a random function, for
which a functional central limit theorem holds; in fact, because $B_{L}(r)$
has independent increments and finite moments of all order, it is not
difficult to show that, as $L\rightarrow \infty $, 
\begin{equation}
B_{L}(r)\Rightarrow B(r)  \label{conv}
\end{equation}%
where $\Rightarrow $ denotes convergence in distribution in a functional
sense (see for instance \cite{bill}): this ensures, for instance, that
the distribution of functionals of $B_{L}(r)$ will converge to the
distribution of the same functional, evaluated on $B(r).$ Also, $B(r)$
denotes the well known standard Brownian motion process, whose properties
are widely studied and well known: it is a Gaussian, zero-mean continuous
process, with independent increments such that 
\begin{equation}
B(r_2) - B(r_1) \stackrel{d}{=} N(0,r_2 - r_1) \mbox{ .}
\end{equation}%
In view of equation~\ref{conv} and standard properties of Brownian motion, we are
for instance able to conclude that 
\begin{equation}
\fl \lim_{L\rightarrow \infty }P\left\{ \sup_{r\in \lbrack
0,1]}B_{L}(r)>x\right\} =P\left\{ \sup_{r\in \lbrack 0,1]}B(r)>x\right\}
=2P(Z>x),   \label{pol1}
\end{equation}%
$Z$ denoting a standard (zero mean, unit-variance) Gaussian variable (see
for instance \cite{borsan}). This means that to determine
approximate threshold values for the maximum value of the sum $\sum_{\ell
=1}^{[Lr]}H_{\ell }$ as $r$ varies between zero and one, the tables of a
standard Gaussian variate are sufficient. Likewise, the asymptotic
distribution of $P\left\{ \sup_{r}|B_{L}(r)|>x\right\} $ is given by%
\begin{eqnarray} \label{pol2}
\fl \lim_{L\rightarrow \infty }P\left\{ \sup_{r}|B_{L}(r)|>x\right\} = \\
\fl = \frac{1}{%
\sqrt{2\pi }}\sum_{k=-\infty }^{\infty }\int_{-x}^{x}\left[\exp \left(
  -\frac{%
(z+4kx)^{2}}{2}\right)-\exp \left(-\frac{(z+2x+4kx)^{2}}{2}\right)\right]dz \mbox{ .} \nonumber
\end{eqnarray}%
Monte Carlo simulations have confirmed that equation~\ref{pol1} and equation~\ref{pol2}
provide accurate approximations of the finite sample distributions, for $L$
in the order of $10^{3}$.

\section{Effect of noise correlation}

In order to consider the effect of correlated noise we start discussing
the simplest case with two detectors. The presence of correlated noise
can be inserted by rewriting equation~\ref{eq:dl} as
\begin{eqnarray} \label{eq:dlcorr}
  d_{\ell m}^{(1)}&=&a_{\ell m}+a_{\ell m}^{N_1}+c_{\ell m}^{(1 2)} \\
  \nonumber
  d_{\ell m}^{(2)}&=&a_{\ell m}+a_{\ell m}^{N_2}+c_{\ell m}^{(1 2)}
\end{eqnarray}
where $c_{\ell m}^{(1 2)}$ is independent from $a_{\ell m}$, $a_{\ell m}^{N_1}$
and $a_{\ell m}^{N_2}$.
Under these circumstances,
it is clear that both ${\widehat{C}}_{\ell}$ and ${\widetilde{C}}_{\ell}$ will be biased; however, their
difference $G_{\ell}$, used in the Hausman test, is not affected at
all due to cancellations of all the terms involving $c_{\ell m}^{(1 2)}$:
\begin{eqnarray}
\fl G_{l} &=&  \left\{ {\widehat{C}}_{\ell} - {\widetilde{C}}_{\ell}
\right\} =
 \frac{1}{2\ell +1} \sum_{m=-\ell}^{\ell} \left\{
\left(
|a_{\ell m}|^2 + a_{\ell m}(\overline{a}_{\ell m}^{N_1} +
\overline{a}_{\ell m}^{N_2} + 2 \overline{c}_{\ell  m}^{(1 2)})+
\right. \right. \\
 \fl && \left. \frac{1}{4}(|{a}_{\ell m}^{N_1}|^2+ |{a}_{\ell m}^{N_2}|^2+
 2 {a}_{\ell m}^{N_1} \overline{a}_{\ell m}^{N_2}) +
 ({a}_{\ell m}^{N_1} + {a}_{\ell m}^{N_2}) \overline{c}_{\ell m}^{(1 2)} +  
 |{c}_{\ell m}^{(1 2)}|^2
 \right) - \nonumber \\
 \fl &&\left. \left(  |a_{\ell m}|^2 + a_{\ell m}(\overline{a}_{\ell m}^{N_1} + 
 \overline{a}_{\ell m}^{N_2} + 2 \overline{c}_{\ell m}^{(1 2)}) + 
 {a}_{\ell m}^{N_1} \overline{a}_{\ell m}^{N_2} +
 ({a}_{\ell m}^{N_1} + {a}_{\ell m}^{N_2})\overline{c}_{\ell m}^{(1 2)} +  
 |{c}_{\ell m}^{(1 2)}|^2 \right)
 \right\} \nonumber \\
\fl &=&\frac{1}{2\ell +1} \sum_{m=-\ell}^{\ell} \frac{1}{4} \left( 
|{a}_{\ell m}^{N_1}|^2+ |{a}_{\ell m}^{N_2}|^2 - 2 {a}_{\ell m}^{N_1} \overline{a}_{\ell m}^{N_2}
\right) \mbox{ .} \nonumber 
\end{eqnarray}

In the more general case with $k$ detectors with noise correlations
which varies from pair to pair, this is no longer true.
In fact, by completely analogous arguments, it can be shown that some extra
terms involving cross-products of the form
 ${a}_{\ell m}^{N_i} {c}_{\ell m}^{i j}$ will remain in $G_{\ell}$.
These terms, however, have zero expected value, and thus
will affect only the variance of $H_{\ell}$.
In other words, the previous approach can go through unaltered,
provided we have available a reliable estimate of the variance of
$G_{\ell}$.
These issues are investigated by means of Monte Carlo simulations in the next section.

\section{Monte Carlo simulations}

\label{sec:mc}
\begin{figure}[htp]
\begin{center}
\includegraphics[scale=0.6]{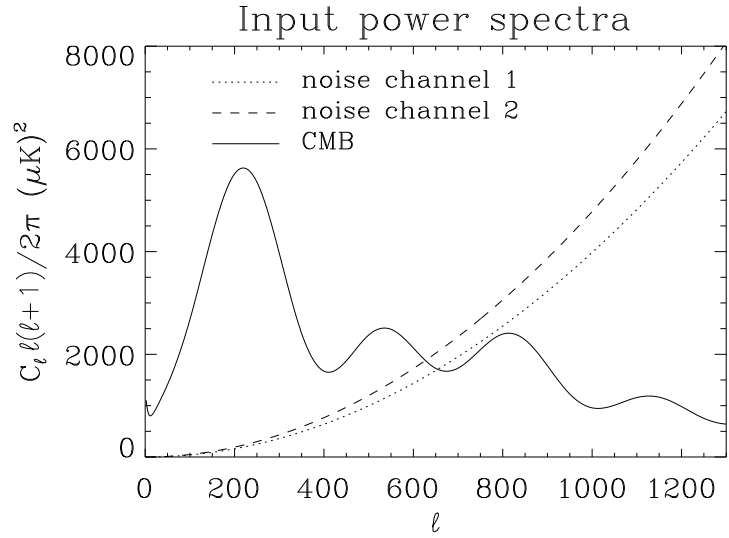}
\end{center}
\caption{\label{fig:inputps} Input power spectra used in the simulation. The RMS of the noise
per $7'$ pixel is $55$ and $65 \mu K$ for channel 1 and channel 2 respectively.
}
\end{figure}

To verify the validity of the previous analytic arguments, we
present in this section some Monte Carlo simulations. As a first step, we
generate some Gaussian, full sky CMB maps from a parent distribution with a
given power spectrum, which corresponds to a standard $\Lambda CDM$ model
with running spectral index; the values of the parameters are provided by
the WMAP best fit, that is $\Omega _{b}h^{2}=0.02262,$ $\Omega
_{CDM}h^{2}=0.10861,$ $n(k=0.05Mpc^{-1})=1.04173,$ $exp(-2\tau )=0.69879,$ $dn/d\ln
k=-0.01618,$ $amp(k=0.05Mpc^{-1})=0.86746,$ $h=0.73070.$
In order to include the effect of a finite resolution of the detectors, we
simulate the maps using a beam of $12'$ FWHM.
Then we considered two channels
and added random Gaussian noises realizations to each of them;
noise is assumed to be white and isotropic with RMS amplitude per $7'$ pixels of 
$55 \mu K$ and $65 \mu K$ respectively for the two channels. 
The input power spectra used are 
shown if figure~\ref{fig:inputps}.
We start considering full sky maps.
From each CMB realization we compute both the cross-power
spectrum and the auto-power spectrum, for $l=2,...,L=1300$. We generated
1000 maps, and we start by presenting the Monte Carlo values for the
variances of the cross-spectrum and auto-power spectrum estimators, together
with the variance of their differences. Results are shown in figures~\ref%
{fig:crvar}-\ref{fig:hausvar}; they are
clearly in extremely good agreement with the values that were obtained
analytically.

\begin{figure}[htp]
\begin{center}
\includegraphics[scale=0.6]{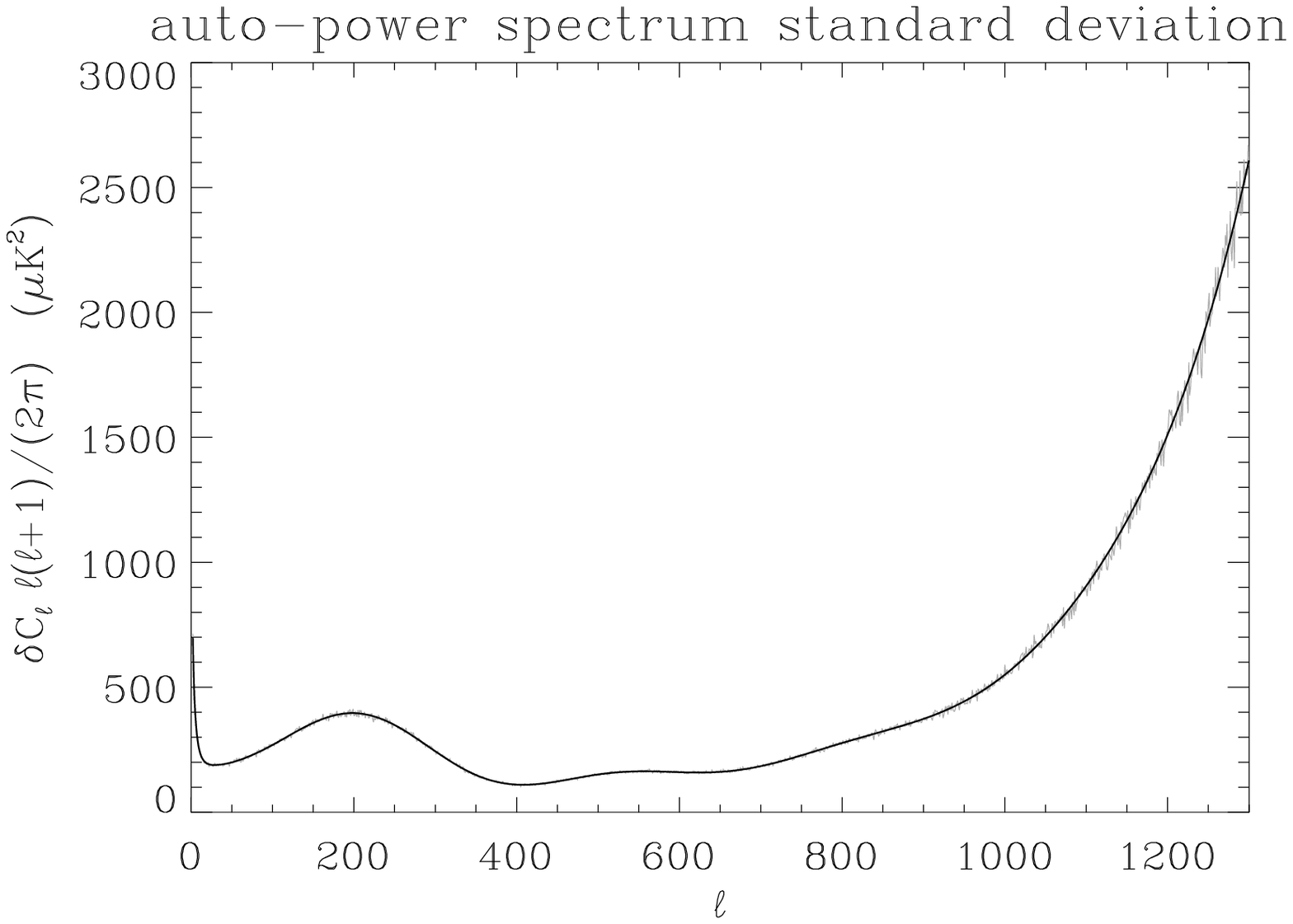}
\end{center}
\caption{\label{fig:crvar} Standard deviation of the auto-power spectrum estimator. In grey
we show the results obtained from 1000 Monte Carlo simulations, while the
black line is obtained from equation~\ref{eq:final}.}
\end{figure}

\begin{figure}[htp]
\begin{center}
\includegraphics[scale=0.6]{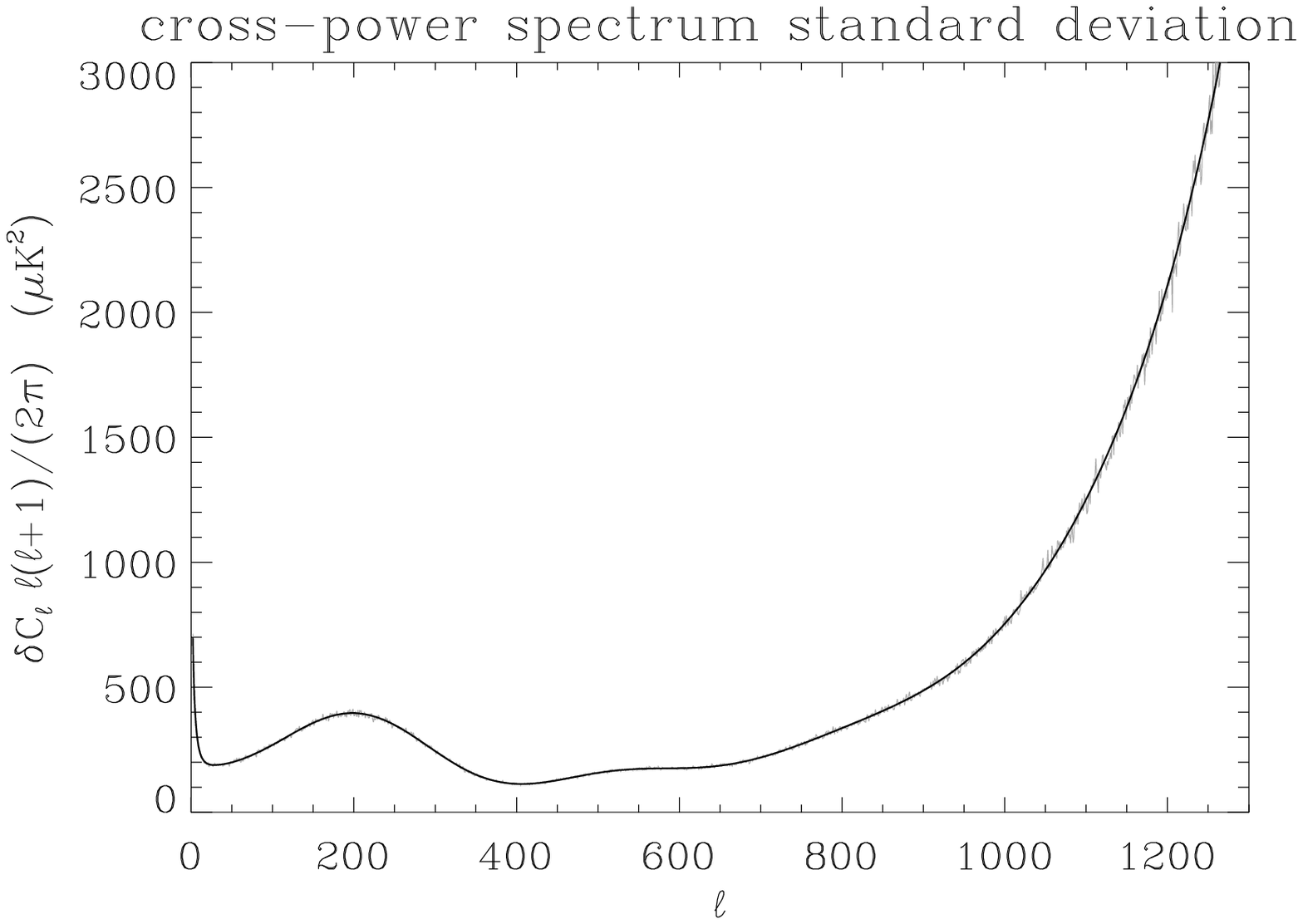}
\end{center}
\caption{\label{fig:mvar}Standard deviation of the cross-power spectrum estimator. In grey
we show the results obtained from 1000 Monte Carlo simulations, while the
black line is obtained from equation~\ref{mastervar}.}
\end{figure}

\begin{figure}[tph]
\begin{center}
\includegraphics[scale=0.6]{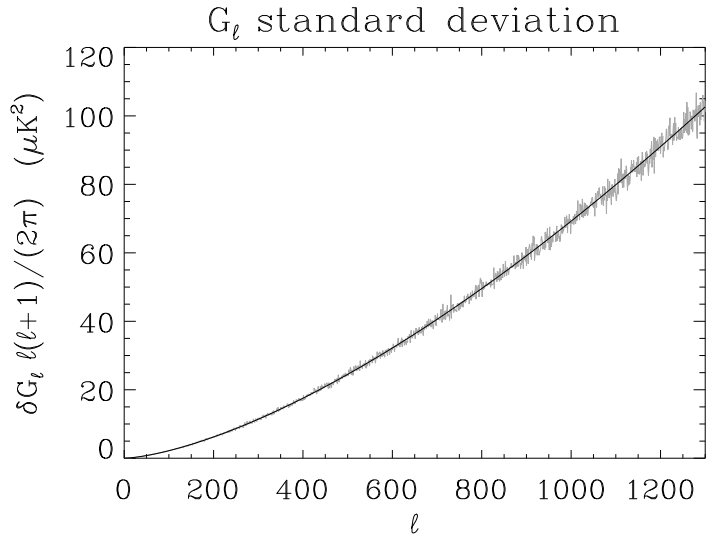}
\end{center}
\caption{\label{fig:hausvar}Standard deviation of the difference between the auto- and the
cross-power spectrum estimator. In grey we show the results obtained from
1000 Monte Carlo simulations, while the black line is obtained from equation~\ref%
{eq:hausvar}.}
\end{figure}
We now focus more directly on the efficiency of the Hausman test in
identifying a residual bias in the auto-power spectrum. In order to achieve
this goal,
we simulate 300 further maps with a noise power spectrum ${C}_{\ell }^{N}$, and
we compute the auto-power spectrum
using a modified version of equation~\ref{basest}: 
\begin{equation}
\widehat{C}_{\ell }=\frac{1}{2\ell +1}\sum_{m=-\ell }^{\ell }d_{\ell m}%
\overline{d_{\ell m}}-f_{n} C_{\ell }^{N}  \mbox{ .}
\end{equation}%
In this way we simulate a wrong estimation of the noise power spectrum.

 Then, for a fixed $f_{n}$, we compute $H_{\ell }$ and $B_{L}(r)$ for each simulation. We
consider the three test statistics $s_{1}=\sup_{r}B_{L}(r)$, $%
s_{2}=\sup_{r}|B_{L}(r)|$ and $s_{3}=\int_{0}^{1}B_{L}^{2}(r)dr$, and the
threshold values for the $68\%,95\%$ and $99\%$ probability. We used a
thousand independent simulations with the value $f_{n}=1,$, corresponding to
the case where our {\it a priori} knowledge of noise is correct, to tabulate the
empirical distributions under this null hypothesis; results are reported in table~\ref{tab:thres}.

\begin{table}
\caption{\label{tab:thres} Threshold values under the null hypothesis }
\begin{indented}
\item[]\begin{tabular}{@{}llll}
\br
& $s_{1}$ & $s_{2}$ & $s_{3}$ \\ 
\mr
$68\%$ & 0.919 & 1.533 & 0.596 \\ 
$95\%$ & 1.818 & 2.550 & 1.926 \\ 
$99\%$ & 2.401 & 3.308 & 3.842 \\%
\br
\end{tabular}
\end{indented}
\end{table}

We then go on to compute $s_{1},s_{2}$ and $s_{3}$ under the alternatives $%
f_{n}\neq 1;$ the percentages of rejections provide an estimate of the power
of these procedures in detecting a bias. Results are reported in tables~\ref{tab:s1}-\ref{tab:s3},
and are clearly very encouraging: the $s_{2}$ and $s_{3}$ test statistics
enjoy $100\%$ power
even in the presence of a mere $1\%$ misspecification of the noise angular
power spectrum. Note that, as expected, $s_{1}$ is a unidirectional test,
that is, it has no power in the case where noise is overestimated ($f_{n}>1)$;
however, for such circumstances it would suffice to consider $s_{1}^{\prime
}=\inf_{r}B_{L}(r)$ to obtain satisfactory power properties. In general, $%
s_{2}$ and $s_{3}$ should clearly be preferred for their robustness against
a wider class of departures from the null.

\begin{table}[ht]
\caption{\label{tab:s1}  The power of the test $s_{1}$}
\begin{indented}
\lineup
\item[]\begin{tabular}{@{}lllllllll}
\br
$f_{n}$ & 0.99 & 0.995 & 0.998 & 0.999 & 1.001 & 1.002 & 1.005 & 1.01 \\ 
\mr
$68\%$ & 1.00 & 1.00 & 0.79 & 0.54 & 0.15 & 0.08 & 0.01 & 0.00 \\ 
$95\%$ & 1.00 & 0.99 & 0.39 & 0.14 & 0.01 & 0.00 & 0.00 & 0.00 \\ 
$99\%$ & 1.00 & 0.96 & 0.16 & 0.05 & 0.00 & 0.00 & 0.00 & 0.00 \\
\br
\end{tabular}
\end{indented}
\end{table}

\begin{table}[ht]
\caption{\label{tab:s2}  The power of the test $s_{2}$}
\begin{indented}

\item[]\begin{tabular}{@{}lllllllll}
\br
$f_{n}$ & 0.99 & 0.995 & 0.998 & 0.999 & 1.001 & 1.002 & 1.005 & 1.01 \\ 
\mr
$68\%$ & 1.00 & 1.00 & 0.54 & 0.27 & 0.53 & 0.82 & 1.00 & 1.00 \\ 
$95\%$ & 1.00 & 0.95 & 0.12 & 0.04 & 0.11 & 0.43 & 0.98 & 1.00 \\ 
$99\%$ & 1.00 & 0.76 & 0.04 & 0.00 & 0.03 & 0.12 & 0.95 & 1.00 \\%
\br
\end{tabular}
\end{indented}
\end{table}

\begin{table}[ht]
\caption{\label{tab:s3}  The power of the test $s_{3}$}
\begin{indented}

\item[]\begin{tabular}{@{}lllllllll}
\br
$f_{n}$ & 0.99 & 0.995 & 0.998 & 0.999 & 1.001 & 1.002 & 1.005 & 1.01 \\ 
\mr
$68\%$ & 1.00 & 0.99 & 0.57 & 0.33 & 0.45 & 0.74 & 1.00 & 1.00 \\ 
$95\%$ & 1.00 & 0.86 & 0.15 & 0.06 & 0.11 & 0.30 & 0.95 & 1.00 \\ 
$99\%$ & 1.00 & 0.58 & 0.03 & 0.01 & 0.02 & 0.08 & 0.76 & 1.00 \\%
\br
\end{tabular}
\end{indented}
\end{table}

\subsection{Effect of partial sky coverage}

In order to study the effect of partial sky coverage on the Hausman
test, we repeated the Monte Carlo analysis considering the patch
observed by BOOMERanG, covering $\sim 2\%$ of the full sky \cite{montroy,masi05}.
We expect this to be a good limiting case,
where any failures of the test due to partial sky coverage should
clearly show up.

The main effect of partial sky coverage is, as well known, to produce correlations
among spherical harmonic coefficients, that can be interpreted as a
reduction of the effective number of degrees of freedom in the power
spectrum \cite{efh01}. 

Results are reported in tables~\ref{tab:s1cut}-\ref{tab:s3cut}.
We stress that the power of the Hausman test, although reduced (as expected), 
 is still very satisfactory. For instance, a misspecification of the
 noise level of the order of $5\%$ is detected $100\%$ of the times by
 $s_2$ and $99\%$ by $s_3$.

\begin{table}[ht]
\caption{\label{tab:s1cut}  The power of the test $s_{1}$ in the
  presence of partial sky coverage.}
\begin{indented}
\lineup
\item[]\begin{tabular}{@{}lllllllllll}
\br
$f_{n}$ & 0.95 & 0.96 & 0.97 & 0.98 & 0.99 & 1.01 & 1.02 & 1.03 & 1.04
& 1.05\\ 
\mr
$68\%$ & 1.00 & 1.00 & 0.99 & 0.94 & 0.70 & 0.09 & 0.03 & 0.01 & 0.00
& 0.00\\ 
$95\%$ & 1.00 & 1.00 & 0.97 & 0.75 & 0.35 & 0.01 & 0.00 & 0.00 & 0.00
& 0.00\\ 
$99\%$ & 1.00 & 0.99 & 0.91 & 0.63 & 0.17 & 0.01 & 0.00 & 0.00 & 0.00
& 0.00\\
\br
\end{tabular}
\end{indented}
\end{table}

\begin{table}[ht]
\caption{\label{tab:s2cut}  The power of the test $s_{2}$ in the
  presence of partial sky coverage.}
\begin{indented}

\item[]\begin{tabular}{@{}lllllllllll}
\br
$f_{n}$ & 0.95 & 0.96 & 0.97 & 0.98 & 0.99 & 1.01 & 1.02 & 1.03 & 1.04
& 1.05\\ 
\mr
$68\%$ & 1.00 & 1.00 & 0.99 & 0.88 & 0.58 & 0.42 & 0.82 & 0.98 & 1.00
& 1.00\\ 
$95\%$ & 1.00 & 1.00 & 0.95 & 0.67 & 0.26 & 0.18 & 0.54 & 0.91 & 0.99
& 1.00\\ 
$99\%$ & 1.00 & 0.99 & 0.85 & 0.50 & 0.11 & 0.08 & 0.34 & 0.78 & 0.97
& 1.00\\
\br
\end{tabular}
\end{indented}
\end{table}

\begin{table}[ht]
\caption{\label{tab:s3cut}  The power of the test $s_{3}$ in the
  presence of partial sky coverage.}
\begin{indented}

\item[]\begin{tabular}{@{}lllllllllll}
\br
$f_{n}$ & 0.95 & 0.96 & 0.97 & 0.98 & 0.99 & 1.01 & 1.02 & 1.03 & 1.04
& 1.05\\ 
\mr
$68\%$ & 1.00 & 1.00 & 0.96 & 0.80 & 0.51 & 0.37 & 0.70 & 0.94 & 0.99
& 1.00\\ 
$95\%$ & 1.00 & 0.96 & 0.82 & 0.53 & 0.16 & 0.13 & 0.35 & 0.73 & 0.95
& 0.99\\ 
$99\%$ & 0.99 & 0.92 & 0.70 & 0.33 & 0.09 & 0.05 & 0.23 & 0.55 & 0.88
& 0.99\\
\br
\end{tabular}
\end{indented}
\end{table}

\subsection{Polarization and 1/f noise correlated among different detectors}

We move forward, analysing a more realistic case including
polarization measurements in the
presence of 1/f noise correlated among different detectors. 
This is achieved generating time ordered data with a scanning strategy
and detectors noise properties
similar to those of BOOMERanG-B03, where correlations of the
order of $10\%$ are present (see table 7 in \cite{masi05}), and the
1/f noise knee frequency is $\sim 0.07Hz$ (see figure 21 in \cite{masi05}). The sky maps are
then obtained using the ROMA IGLS polarization map-making code \cite{deg05}.

Polarization measurements provide six power spectra that can be
used separately or combined to obtain a more efficient detection of the noise bias.
The optimal combination of polarization power spectra is
under investigation and will be addressed in a future paper. 
Here, in order to illustrate the method, we simply average the $B_L(r)$
obtained from each power spectrum.

Results are reported in tables~\ref{tab:s1pol}-\ref{tab:s3pol}.
Once more the power of the Hausman test is reduced with respect to the
full sky uncorrelated noise case, but is still satisfactory. For instance, a misspecification of the
noise level of the order of $15\%$ is detected $\sim 100\%$ of the
times with $95\%$ significance.

\begin{table}[ht]
\caption{\label{tab:s1pol}  The power of the test $s_{1}$ for polarization measurements with 1/f noise correlated among different detectors.}
\begin{indented}
\lineup
\item[]\begin{tabular}{@{}lllllllllll}
\br
$f_{n}$ & 0.80 & 0.85 & 0.90 & 0.95 & 0.98 & 1.02 & 1.05 & 1.10 & 1.15
& 1.20\\ 
\mr
$68\%$ & 1.00 & 1.00 & 1.00 & 0.93 & 0.79 & 0.48 & 0.29 & 0.12 & 0.05
& 0.00\\ 
$95\%$ & 1.00 & 1.00 & 0.89 & 0.47 & 0.21 & 0.06 & 0.04 & 0.02 & 0.01
& 0.00\\ 
$99\%$ & 0.99 & 0.88 & 0.45 & 0.11 & 0.05 & 0.02 & 0.01 & 0.00 & 0.00
& 0.00\\
\br
\end{tabular}
\end{indented}
\end{table}

\begin{table}[ht]
\caption{\label{tab:s2pol}  The power of the test $s_{2}$ for polarization measurements with 1/f noise correlated among different detectors.}
\begin{indented}

\item[]\begin{tabular}{@{}lllllllllll}
\br
$f_{n}$ & 0.80 & 0.85 & 0.90 & 0.95 & 0.98 & 1.02 & 1.05 & 1.10 & 1.15
& 1.20\\ 
\mr
$68\%$ & 1.00 & 1.00 & 0.99 & 0.83 & 0.65 & 0.76 & 0.90 & 1.00 & 1.00
& 1.00\\ 
$95\%$ & 1.00 & 0.99 & 0.73 & 0.29 & 0.12 & 0.17 & 0.39 & 0.86 & 0.99
& 1.00\\ 
$99\%$ & 0.99 & 0.83 & 0.38 & 0.07 & 0.04 & 0.04 & 0.07 & 0.51 & 0.93
& 1.00\\
\br
\end{tabular}
\end{indented}
\end{table}

\begin{table}[ht]
\caption{\label{tab:s3pol}  The power of the test $s_{3}$ for polarization measurements with 1/f noise correlated among different detectors.}
\begin{indented}

\item[]\begin{tabular}{@{}lllllllllll}
\br
$f_{n}$ & 0.80 & 0.85 & 0.90 & 0.95 & 0.98 & 1.02 & 1.05 & 1.10 & 1.15
& 1.20\\ 
\mr
$68\%$ & 1.00 & 1.00 & 0.98 & 0.84 & 0.72 & 0.75 & 0.89 & 1.00 & 1.00
& 1.00\\ 
$95\%$ & 1.00 & 0.95 & 0.61 & 0.21 & 0.14 & 0.17 & 0.36 & 0.76 & 0.96
& 1.00\\ 
$99\%$ & 0.97 & 0.70 & 0.25 & 0.09 & 0.04 & 0.05 & 0.07 & 0.36 & 0.82
& 0.99\\
\br
\end{tabular}
\end{indented}
\end{table}

The Planck experiment will provide full sky polarization maps with
pixel sensitivity similar to that of BOOMERanG-B03. Such a wide sky
coverage will allow us to reach unprecedented accuracy in the estimated
power spectra. The application of the Hausman test to Planck simulated
maps will be the subject of a forthcoming paper.

\section{Conclusions}

\label{sec:con} We have discussed the analytic properties of the cross-power
spectrum as an estimator of the angular power spectrum of the CMB
anisotropies. The method is computationally convenient for very large data
sets as those provided by WMAP or Planck and it provides unbiased estimates
under very broad assumptions (basically, that noise is uncorrelated along
different channels).
It thus provides a robust alternative, where noise
estimation and subtraction are not required. We also propose a new procedure
for testing for the presence of residual bias due to inappropriate noise subtraction
in pseudo-$C_{\ell}$ estimates (the Hausman test). The test compares
the auto- and cross-power spectrum estimators under the null
hypothesis. In the case of failure, the more robust cross-power spectrum
should be preferred, while in the case of success both estimators could be
used, and the choice should result from a trade-off between efficiency
and robustness.
We derive the analytic behaviour of this procedure
under the null hypothesis, and use Monte Carlo simulations to investigate its power
properties, which appear extremely promising. We leave for future research
some further improvements of this approach, in particular, 
the use of bootstrap/resampling methods to make
even the determination of confidence intervals independent from noise
estimation. Finally, the optimal combination of polarization power
spectra and the application of the Hausman test to the Planck satellite are currently
under investigation.

\ack
The authors would like to thank G. De Gasperis for providing
realistic temperature and polarization simulated maps and F.K. Hansen,
P. Cabella, G. De Troia, F. Piacentini
and K. Ganga for useful discussions. We acknowledge the use of the
MASTER, HEALPix, CMBFAST and FFTW packages. Research supported by MURST, ASI, PNRA.

\appendix
\section*{Appendix}
\label{sec:ap}
\setcounter{section}{1}

We recall the definition of the cross spectrum estimator (equation~\ref{eq:def}): 
\begin{eqnarray}
\fl \widetilde{C}_{\ell }^{ij} =\frac{1}{2\ell +1}\sum_{m=-\ell }^{\ell
}d_{\ell m}^{i}\overline{d_{\ell m}^{j}} =\frac{1}{2\ell
  +1}\sum_{m=-\ell }^{\ell }\left\{ (a_{\ell m}+a_{\ell m}^{N_{i}})(%
\overline{a}_{\ell m}+\overline{a}_{\ell m}^{N_{j}})\right\} \nonumber \\
\fl = \frac{1}{2\ell +1}\sum_{m=1}^{\ell }\left\{ (2|a_{\ell
    m}|^{2}+a_{\ell m}^{N_{i}}%
\overline{a}_{\ell m}+\overline{a}_{\ell m}^{N_{i}}a_{\ell m}+a_{\ell m}\overline{a}%
_{\ell m}^{N_{j}}+\overline{a}_{\ell m}a_{\ell m}^{N_{j}}+a_{\ell m}^{N_{i}}\overline{a}%
_{\ell m}^{N_{j}}+\overline{a}_{\ell m}^{N_{i}}a_{\ell m}^{N_{j}})\right\}   \nonumber
\\
+\frac{1}{2\ell +1}\left\{
(|a_{\ell 0}|^{2}+a_{\ell 0}^{N_{i}}a_{\ell 0}+a_{\ell 0}a_{\ell
  0}^{N_{j}}+a_{\ell m}^{N_{i}}a_{\ell m}^{N_{j}})\right\} \mbox{ .}
\label{due}
\end{eqnarray}%
It is easy to see that all summands in equation~\ref{due} are
uncorrelated (albeit not independent), with variances given by%
\begin{eqnarray}
Var\left\{ 2|a_{\ell m}|^{2}\right\}  = 4C_{\ell }, m=1,...,\ell
, Var\left\{ |a_{\ell 0}|^{2}\right\} =2C_{\ell } \\
Var\left\{ a_{\ell m}^{N_{i}}\overline{a}_{\ell m}\right\}  = Var\left\{ \overline{a%
}_{\ell m}^{N_{i}}a_{\ell m}\right\} =C_{\ell }^{N_{i}}C_{\ell }, %
m=0,...,\ell  \\
Var\left\{ a_{\ell m}^{N_{i}}\overline{a}_{\ell m}^{N_{j}}\right\}  =Var\left\{ 
\overline{a}_{\ell m}^{N_{i}}a_{\ell m}^{N_{j}}\right\} =C_{\ell }^{N_{i}}C_{\ell
}^{N_{j}}, m=0,...,\ell 
\end{eqnarray}%
whence we obtain 
\begin{eqnarray}
\fl Var\left\{ \widetilde{C}_{\ell }^{ij}\right\}  =\frac{2}{(2\ell +1)^{2}}%
\sum_{m=1}^{\ell }\left\{ 2C_{\ell }+C_{\ell }^{N_{i}}C_{\ell }+C_{\ell
}^{N_{j}}C_{\ell }+C_{\ell }^{N_{i}}C_{\ell }^{N_{j}}\right\}   \nonumber \\
 +\frac{1}{(2\ell +1)^{2}}\left\{ 2C_{\ell }+C_{\ell }^{N_{i}}C_{\ell
}+C_{\ell }^{N_{j}}C_{\ell }+C_{\ell }^{N_{i}}C_{\ell }^{N_{j}}\right\}  
\nonumber \\
\lo{=}\frac{1}{2\ell +1}\left\{ 2C_{\ell }+C_{\ell }^{N_{i}}C_{\ell }+C_{\ell
}^{N_{j}}C_{\ell }+C_{\ell }^{N_{i}}C_{\ell }^{N_{j}}\right\}  \mbox{ .}
\end{eqnarray}%
\newline
We show now that, for a single couple $(i,j),$ we have 
\begin{equation}
\fl Cov\left\{ \widehat{C}_{\ell },\widetilde{C}_{\ell }^{ij}\right\} =\frac{2}{%
2\ell +1}\left\{ C_{\ell }^{2}+\frac{C_{\ell }}{k}\left( C_{\ell
}^{N_{i}}+C_{\ell }^{N_{j}}\right) +\frac{1}{k^{2}}C_{\ell }^{N_{i}}C_{\ell
}^{N_{j}}\right\}   \mbox{ .}
\end{equation}%
Indeed,%
\begin{eqnarray}
\fl Cov\left\{ \widehat{C}_{\ell },\widetilde{C}_{\ell }^{ij}\right\}
= \nonumber \\
\fl = Cov\left[ \left\{ |a_{\ell m}|^{2}+a_{\ell m}\overline{a}_{\ell m}^{N}+%
\overline{a}_{\ell m}a_{\ell m}^{N}+|a_{\ell m}^{N}|^{2}\right\} ,\left\{
|a_{\ell m}|^{2}+a_{\ell m}\overline{a}_{\ell m}^{N_{j}}+a_{\ell m}^{N_{i}}%
\overline{a}_{\ell m}+a_{\ell m}^{N_{i}}\overline{a}_{\ell
m}^{N_{j}}\right\} \right] 
\nonumber \\
\fl =\frac{1}{(2\ell +1)^{2}}Cov\left[ \left\{ a_{\ell 0}^{2}+2a_{\ell 0}a_{\ell
0}^{N}+(a_{\ell 0}^{N})^{2}\right\} ,\left\{ a_{\ell 0}^{2}+a_{\ell
0}a_{\ell 0}^{N_{j}}+a_{\ell 0}^{N_{i}}a_{\ell 0}+a_{\ell 0}^{N_{i}}a_{\ell
0}^{N_{j}}\right\} \right] \nonumber \\
 +\frac{1}{(2\ell +1)^{2}}\sum_{m=1}^{\ell }Cov\left[ 2\left\{ |a_{\ell
m}|^{2}+a_{\ell m}\overline{a}_{\ell m}^{N}+\overline{a}_{\ell m}a_{\ell
m}^{N}+|a_{\ell m}^{N}|^{2}\right\} ,\right.  \\
 \left. \left\{ 2|a_{\ell m}|^{2}+a_{\ell m}\overline{a}_{\ell m}^{N_{j}}+%
\overline{a}_{\ell m}a_{\ell m}^{N_{J}}+a_{\ell m}\overline{a}_{\ell
m}^{N_{i}}+\overline{a}_{\ell m}a_{\ell m}^{N_{i}}+a_{\ell m}^{N_{I}}%
\overline{a}_{\ell m}^{N_{j}}+\overline{a}_{\ell m}^{N_{I}}a_{\ell
m}^{N_{j}}\right\} \right]   \nonumber
\end{eqnarray}%
where%
\begin{equation}
a_{\ell m}^{N}=\frac{1}{k}\sum_{i=1}^{k}a_{\ell m}^{N_{i}}  \mbox{ .}
\end{equation}%
Now,%
\begin{eqnarray}
\fl Cov\left[ \left\{ a_{\ell 0}^{2}+2a_{\ell 0}a_{\ell 0}^{N}+(a_{\ell
0}^{N})^{2}\right\} ,\left\{ a_{\ell 0}^{2}+a_{\ell 0}a_{\ell
0}^{N_{j}}+a_{\ell 0}^{N_{i}}a_{\ell 0}+a_{\ell 0}^{N_{i}}a_{\ell
0}^{N_{j}}\right\} \right]  \\
\lo{=} 2\left\{ C_{\ell }^{2}+\frac{C_{\ell }}{k}(C_{\ell
}^{N_{i}}+C_{\ell }^{N_{j}})+\frac{1}{k^{2}}C_{\ell }^{N_{i}}C_{\ell
}^{N_{j}}\right\} \nonumber
\end{eqnarray}%
and likewise%
\begin{eqnarray}
\fl Cov\left[ 2\left\{ |a_{\ell m}|^{2}+a_{\ell m}\overline{a}_{\ell m}^{N}+%
\overline{a}_{\ell m}a_{\ell m}^{N}+|a_{\ell m}^{N}|^{2}\right\} ,\right.  
\nonumber \\
 \left. \left\{ 2|a_{\ell m}|^{2}+a_{\ell m}\overline{a}_{\ell m}^{N_{j}}+%
\overline{a}_{\ell m}a_{\ell m}^{N_{J}}+a_{\ell m}\overline{a}_{\ell
m}^{N_{i}}+\overline{a}_{\ell m}a_{\ell m}^{N_{i}}+a_{\ell m}^{N_{I}}%
\overline{a}_{\ell m}^{N_{j}}+\overline{a}_{\ell m}^{N_{I}}a_{\ell
m}^{N_{j}}\right\} \right]   \nonumber \\
 \lo{=} 4\left\{ C_{\ell }^{2}+\frac{C_{\ell }}{k}(C_{\ell }^{N_{i}}+C_{\ell
}^{N_{j}})+\frac{1}{k^{2}}C_{\ell }^{N_{i}}C_{\ell }^{N_{j}}\right\}  \mbox{ .}
\end{eqnarray}%
Hence, 
\begin{eqnarray}
\fl Cov\left\{ \widehat{C}_{\ell },\widetilde{C}_{\ell }^{ij}\right\} =\frac{2%
}{(2\ell +1)^{2}}\left\{ C_{\ell }^{2}+\frac{C_{\ell }}{k}(C_{\ell
}^{N_{i}}+C_{\ell }^{N_{j}})+\frac{1}{k^{2}}C_{\ell }^{N_{i}}C_{\ell
}^{N_{j}}\right\}   \nonumber \\
 +\frac{4}{(2\ell +1)^{2}}\sum_{m=1}^{\ell }\left\{ C_{\ell }^{2}+\frac{%
C_{\ell }}{k}(C_{\ell }^{N_{i}}+C_{\ell }^{N_{j}})+\frac{1}{k^{2}}C_{\ell
}^{N_{i}}C_{\ell }^{N_{j}}\right\}   \nonumber \\
\lo{=}\frac{2}{2\ell +1}\left\{ C_{\ell }^{2}+\frac{C_{\ell }}{k}(C_{\ell
}^{N_{i}}+C_{\ell }^{N_{j}})+\frac{1}{k^{2}}C_{\ell }^{N_{i}}C_{\ell
}^{N_{j}}\right\}
\end{eqnarray}%
as claimed.


\section*{References}
\begin{thebibliography}{99}


 \bibitem{smoot92} Smoot G F \etal 1992 {\it ApJ} {\bf 396} L1
 \bibitem{hanany00} Hanany S \etal 2000 {\it ApJ} {\bf 545} L5
 \bibitem{debe00} de Bernardis Pc\etal 2000 {\it Nature} {\bf 404} 955
 \bibitem{miller99} Miller A D \etal 1999 {\it ApJ} {\bf 524} L1
 \bibitem{dasi02} Halverson N W \etal 2002 {\it ApJ} {\bf 568} 38
 \bibitem{nett02} Netterfield C B \etal 2002 {\it ApJ} {\bf 571} 604
 \bibitem{debe02} de Bernardis P \etal 2002 {\it ApJ} {\bf 564} 559
 \bibitem{ruhl03} Ruhl J \etal 2003 {\it ApJ} {\bf 599} 786
 \bibitem{masi05} Masi S \etal 2005 {\it Preprint} astro-ph/0507509
 \bibitem{jones05} Jones W C \etal 2005 {\it Preprint} astro-ph/0507494
 \bibitem{fp05} Piacentini F \etal 2005 {\it Preprint} astro-ph/0507507
 \bibitem{montroy05} Montroy T E \etal 2005 {\it Preprint} astro-ph/0507514
 \bibitem{cmt05} MacTavish C J \etal 2005 {\it Preprint} astro-ph/0507503
 \bibitem{vsa03} Grainge K \etal 2003 {\it MNRAS} {\bf 341} L23
 \bibitem{benoit03}Benoit A \etal 2003 {\it A\&A} {\bf 399} L19
 \bibitem{cbi02}Pearson T J \etal 2003 {\it ApJ} {\bf 591} 556
 \bibitem{acbar02}Kuo C L \etal 2004 {\it ApJ} {\bf 600} 32
 \bibitem{beast03} O'Dwyer I J \etal 2005 {\it ApJS} {\bf 158} 93
 \bibitem{wmap03}Bennett C L \etal 2003 {\it ApJS} {\bf 148} 1
 \bibitem{bjk98}Bond J R , Jaffe A H and Knox L E 1998 {\it
 Phys. Rev.} D {\bf 57} 2117
 \bibitem{borrill99}Borrill J 1999 {\it AIP Conf Proc} {\bf 476} 277B
 \bibitem{efst03}Efstathiou G 2003 {\it Preprint} astro-ph/0307515
 \bibitem{oh99}Oh S P , Spergel D and Hinshaw G 1999 {\it ApJ} {\bf 510} 550
 \bibitem{wgh01}Wandelt B, Hivon E and Gorski K M 2001 {\it Phys. Rev.} D {\bf64h3003W}
 \bibitem{wh01}Wandelt B and Hansen F K 2003 {\it Phys. Rev.} D {\bf 67b3001W}
 \bibitem{ch01}Challinor A \etal 2003 {\it NewAR} {\bf 47} 995C
 \bibitem{sza01}Szapudi I \etal 2001 {\it ApJ} {\bf 561} L11S 
 \bibitem{efh01}Hivon E \etal 2002 {\it ApJ} {\bf 567} 211
 \bibitem{balbi02}Balbi A \etal 2002 {\it A\&A} {\bf 395} 417b
 \bibitem{fh02}Hansen F K, Gorski K M and Hivon E 2002 {\it MNRAS} {\bf 336} 1304h
 \bibitem{dela}Delabrouille J, Cardoso J F and Patanchon G 2003
 {\it MNRAS} {\bf 346} 1089
 \bibitem{wmapps} Hinshaw G \etal 2003 {\it ApJS} {\bf 148} 135
 \bibitem{pol}Polenta G \etal 2002 {\it ApJ} {\bf 572} L27
 \bibitem{wmapng} Komatsu E \etal 2003 {\it ApJS} {\bf 148} 119
 \bibitem{maxng} Santos M G 2003 {\it MNRAS} {\bf 341} 623
 \bibitem{bill} Billingsley P 1968 {\it Convergence of Probability Measures} (J.Wiley)
 \bibitem{borsan} Borodin A and Salminen P 1996 {\it Handbook of
 Brownian Motion: Facts and Formulae} (Birkhauser)
 \bibitem{montroy} Montroy T E \etal 2003 {\it NewAR}
 {\bf 47} 1057
\bibitem{deg05} De Gasperis G \etal 2005 {\it A\&A} {\bf 436} 1159

\end{thebibliography}
\end{document}